\documentclass[a4paper,12pt]{article}
\usepackage{amsmath}
\usepackage{amssymb}

\newtheorem{theorem}{Theorem}

\newtheorem{corollary}{Corollary}

\newtheorem{definition}{Definition}

\newtheorem{lemma}{Lemma}

\newtheorem{proposition}{Proposition}
\newtheorem{remark}{Remark}

\input{tcilatex}
\begin{document}

\begin{center}
\ \ \ \ \ \ \ \ \ \ \ \ \ \ \ \ \ \ \ \ \ \ \ \ \ \ \ \ \ \ \ \ \ \ \ \ \ \ 

\renewcommand{\thefootnote}{\fnsymbol{footnote}}

{\large GENERAL MANIPULABILITY THEOREM FOR A MATCHING MODEL}\bigskip 

\renewcommand{\thefootnote}{\fnsymbol{footnote}}\vspace{0.4in}

{\large Paola B. Manasero and Jorge Oviedo}\footnotetext{%
Instituto de Matem\'{a}tica Aplicada San Luis, IMASL. Universidad Nacional
de San Luis and CONICET. Italia 1556. D5700HHW San Luis. Rep\'{u}blica
Argentina. E-mail: pbmanasero@unsl.edu.ar (P.B. Manasero) and
joviedo@unsl.edu.ar (J. Oviedo).}\vspace{0.2in}

{\large December 16, 2021\vspace{0.3in}}
\end{center}

\noindent \textsc{abstract:} \ In a many-to-many matching model in which
agents' preferences satisfy substitutability and the law of aggregate
demand, we proof the General Manipulability Theorem. We result generalizes
the presented in Sotomayor (1996 and 2012) for the many-to-one model. In
addition, we show General Manipulability Theorem fail when agents'
preferences satisfy only substitutability.\bigskip 

\medskip 

\medskip 


\thispagestyle{empty}

\section{Introduction}

Many-to-many matching models have been useful for studying assignment
problems with the distinctive feature that agents can be divided into two
disjoint subsets: the set of firms and the set of workers. The nature of the
assignment problem consists of matching each agent with a subset of agents
from the other side of the market. Thus, each firm may hire a subset of
workers while each worker may work for a number of different firms.

\emph{Stability} has been considered the main property to be satisfied by
any matching. A matching is called \emph{stable} if all agents have
acceptable partners and there is no unmatched worker-firm pair who both
would prefer to be matched to each other rather than staying with their
current partners. Unfortunately, the set of stable matchings may be empty. 
\emph{Substitutability}\textit{\footnote{%
Hatfield and Kojima (2010) in matching models with contracts introduce a
weaker condition called \emph{bilateral substitutability} and show that this
condition is sufficient for the existence of a stable matching. Also, they
consider a strengthening of the bilateral substitutability condition called 
\emph{unilateral substitutability}. Both conditions reduce to standard \emph{%
substitutability} in matching problems without contracts. See Definitions 3
and 5 in Hatfield and Kojima (2010) for a precise and formal definition of 
\emph{bilateral }and \emph{unilateral substitutability,} respectively.}} is
the weakest condition that has so far been imposed on agents' preferences
under which the existence of stable matchings is guaranteed. An agent has 
\emph{substitutable} preferences if he wants to continue being matched to an
agent on the other side of the market even if other agents become
unavailable.\footnote{%
Kelso and Crawford (1982) were the first to use \emph{substitutability} to
show the existence of stable matchings in a many-to-one model with money.
Roth (1984) shows that, if all agents have \emph{substitutable} preferences,
the set of many-to-many stable matchings is non-empty.}

The \emph{college admissions problem}\ is the name given by Gale and Shapley
(1962) to a many-to-one matching model. Colleges have \emph{responsive}
preferences over students and students have preference over colleges; each
college $c$ has a maximum number of positions to be filled (its \emph{quota} 
$q_{c}$), it ranks individual students and orders subsets of students in a 
\emph{responsive} manner (Roth 1985); namely, to add \textquotedblleft
good\textquotedblright\ students to a set leads to a better set, whereas to
add \textquotedblleft bad\textquotedblright\ students to a set leads to a
worst set. In addition, for any two subsets that differ in only one student,
the college prefers the subset containing the most preferred student. In
this model the set of stable matchings satisfy the following additional
properties: (i) there is a polarization of interests between the two sides
of the market along the set of stable matchings, (ii) the set of unmatched
agents is the same under every stable matching, (iii) the number of workers
assigned to a firm through stable matchings is the same, and (iv) if a firm
does not complete its quota under some stable matching then it is matched to
the same set of workers at any stable matching.\footnote{%
Property (i) is a consequence of the decomposition lemma proved by Gale and
Sotomayor (1985). Properties (ii) and (iii) were proved independently by
Gale and Sotomayor (1985) and Roth (1984). Property (iv) was proved by Roth
(1986).}

The case in which all quotas are equal to one is called the \emph{marriage
problem},\footnote{%
It is the name given to the one-to-one matching model. See Roth and
Sotomayor (1990) for a precise and formal definition of such model.} and is
symmetric between the two sides of the market. The \emph{college admissions
problem with substitutable preferences}\ is the name given by Roth and
Sotomayor (1990) to the most general many-to-one model with ordinal
preferences. Firms are restricted to have \emph{substitutable} preferences
over subsets of workers, while workers may have all possible preferences
over the set of firms. Under this hypothesis Roth and Sotomayor (1990)
showed that the deferred-acceptance algorithms produce either the
firm-optimal stable matching or the worker-optimal stable matching,
depending on whether the firms or the workers make the offers. The firm
(worker)-optimal stable matching is unanimously considered by all firms
(respectively, workers) to be the best among all stable matchings.

The adoption of a specific rule for some matching model induces a strategic
game where the players are the agents of the model, and the strategies are
the preferences they can state. The payoff function is defined by the given
matching rule. Questions on incentives facing agents naturally emerge.

The first important result in this direction is the \textbf{%
Non-Manipulability Theorem} due to Dubins and Freedman (1981) in the
marriage model. These authors proved that: \textquotedblleft \textit{under
the }$X$\textit{-stable matching rule, the agents of the side of the market }%
$X$\textit{\ of any coalition cannot get preferred mates by falsifying their
preferences}\textquotedblright .\footnote{%
Here, we considerer $X$ the set agents of one side of the market.} In
addition, in the college admissions problem, this result is true of the side
of students (Dubins and Freedman, 1981) and Roth (1985) showed this result
is false of the side of college. In a more general preference domain, for
example, in the many-to-one matching model with substitutable and $q$%
-separable preferences the Non-Manipulability Theorem is true of the side of
workers (Mart\'{\i}nez et. al, 2004) and is false of the side of firms
(Roth, 1985). In the many-to-one matching model with substitutable
preferences the Non-Manipulability Theorem is false in both side of the
market (Roth, 1985 and Mart\'{\i}nez et. al, 2004).

The second important result is the \textbf{General Manipulability Theorem:
\textquotedblleft }\textit{If the matching produced by the allocation rule
is not the optimal stable matching for one of the sides of the market, it is
always true that some participant of this side of the market can be better
off by misrepresent his/her/its preferences}\textquotedblright . In the
college admission problem this result was prove by Sotomayor (1996) of the
side of students and Sotomayor (2012) of the side of college.

The third important result is a \textbf{General Impossibility Theorem} for
the college admission model: \textquotedblleft \textit{Under any stable
matching rule for a given college admission problem, in which there is more
than one stable matching, at least one agent can profitably misrepresent
his/her/its preferences, assuming the others tell the truth}%
\textquotedblright . This result was prove by Sotomayor (2012).

Another important result is the \textbf{Impossibility Theorem: }\textit{%
\textquotedblleft No stable matching rule for the general matching problem
exists for which truthful revelation of preferences is a dominant strategy
for all agents\textquotedblright . In the marriage model }this result was
prove by Roth (1982) and in the college admission was prove by Roth (1985).

The Non-Manipulability Theorem and the General Manipulability Theorem are
the central results of the theory on incentives for the matching model. They
gave origin to or motivated all important results of the theory of stable
matching rules for those models.

Note that if General Manipulability Theorem is true, then General
Impossibility Theorem and Impossibility Theorem holds.

The strict inclusion relationships between the preference domains ($q$%
-responsive\footnote{%
The college admission model is a many-to-one matching model with $q$%
-responsive preferences.} implies $q$-separability and $q$-separability
implies law of aggregated demand), and the absence of inclusion relationship
with respect to substitutability. The domain of $q$-separable and
substitutable preferences is much richer than the domain of responsive
preferences. For instance, consider the case where larger coalitions are
preferred to smaller coalitions (in terms of cardinality). Then,
responsiveness imposes a substantial number of restrictions on the ranking
of coalitions of the same size, whereas separability (together with
substitutability) does not impose any restriction at all.

Hence, in this work, we generalize the General Manipulability Theorem to a
many-to-many matching model such that agents' preferences satisfy
substitutability and law of aggregated demand. In addition, we show that
General Manipulability Theorem fail when the agents' preferences satisfy
only substitutability.

The paper is organized as follow: In Section 2, we present the model. In
Section 3, we present the preliminaries of matching game and state some
results, already proved in the literature, which will be needed in Section
4. In Section 4, we present the manipulability property and the main result.
In Section 5, we present two example show the main result fail when the
model only have substitutable preferences.

\section{The Model}

There are two finite and disjoint sets of agents, the set of $n$ firms $%
F=\left\{ f_{1},...,f_{n}\right\} $ and the set of $m$ workers $W=\left\{
w_{1},...,w_{m}\right\} $. To simplify the notation, sometimes, we denote by 
$f$ (instead of $f_{i}$) whichever firm in $F.$ Similarly, we denote by $w$
(instead of $w_{j}$) whichever worker in $W.$ Each firm $f\in F$ has a
strict linear ordering $\succ _{f}$ over $2^{W}$. And each worker $w\in W$
has a strict linear ordering over $2^{F}$. \textit{Preferences profiles} are 
$\left( n+m\right) $-tuples of preference relations and they are represented
by $\succ =(\succ _{f_{1}},\ldots ,\succ _{f_{n}},\succ _{w_{1}},\ldots
,\succ _{w_{m}})=((\succ _{f})_{f\in F},(\succ _{w})_{w\in W})$.

We denote by $a\in F\cup W$ a generic agent of either set. Given a
preference relation of an agent $\succ _{a},$ the subsets of partners
preferred to the empty set by $a$ are called \emph{acceptable}.

To express preference relations in a concise manner, and since only
acceptable sets of partners will matter, we will represent preference
relations as lists of acceptable partners. For instance, $\succ
_{f_{i}}:\{w_{1},w_{3}\},\{w_{2}\},\{w_{1}\},\{w_{3}\}$ and $\succ
_{w_{j}}:\{f_{1},f_{3}\},\{f_{1}\},\{f_{3}\},$ indicate that $%
\{w_{1},w_{3}\}\succ _{f_{i}}\{w_{2}\}\succ _{f_{i}}\{w_{1}\}\succ
_{f_{i}}\{w_{3}\}\succ _{f_{i}}\emptyset $ and $\{f_{1},f_{3}\}\succ
_{w_{j}}\{f_{1}\}\succ _{w_{j}}\{f_{3}\}\succ _{w_{j}}\emptyset $.

The assignment problem consists of matching workers with firms keeping the
bilateral nature of their relationship and allowing for the possibility that
both, firms and workers, may remain unmatched. Formally,

\begin{definition}
A \textbf{matching} $\mu $ is a mapping from the set $F\cup W$ into the set
of all subsets of $F\cup W$ such that, for all $w\in W$ and $f\in F$:

\begin{enumerate}
\item $\mu (f)\in 2^{W}$\textit{.}

\item $\mu (w)\in 2^{F}$\textit{.}

\item $w\in \mu (f)$ \textit{if and only if }$f\in \mu (w).$\footnote{%
We will often abuse notation by omitting the brackets to denote a set with a
unique element. For instance here, we write $w\in \mu (f)$ instead of $%
\left\{ w\right\} \in \mu (f).$}
\end{enumerate}
\end{definition}

We say that an agent $a$ is \textit{single} in a matching $\mu $ if $\mu
\left( a\right) =\emptyset $. Otherwise, the agent is matched. A matching $%
\mu $ is said to be \textit{one-to-one} if firms can hire at most one
worker, and workers can work for at most one firm. The model in which all
matchings are one-to-one is also known in the literature as the \emph{%
marriage model}. A matching $%
\mu
$ is said to be many-to-one if workers can work for at most one firm but
firms may hire many workers.\bigskip

Given a set of workers $S\subseteq W$, each firm $f\in F$ can determine
which subset of $S$ would most prefer to hire. We will call this $f$'s \emph{%
choice set} from $S$, and denote it by $Ch\left( S,\succ _{f}\right) $.
Formally: 
\begin{equation*}
Ch(S,\succ _{f})=\max_{\succ _{f}}\left\{ T:T\subseteq S\right\}
\end{equation*}%
Symmetrically, given a set of firms $S\subseteq F$ for each worker $w\in W,$
we define: 
\begin{equation*}
Ch(S,\succ _{w})=\max_{\succ _{w}}\left\{ T:T\subseteq S\right\}
\end{equation*}%
A matching $\mu $ is \emph{blocked by agent} $a$ if $\mu \left( a\right)
\neq Ch\left( \mu \left( a\right) ,\succ _{a}\right) $. We say that a
matching is \emph{individually rational} if it is not blocked by any
individual agent.\footnote{%
In a many-to-one model, a matching $\mu $ is individually rational (at $%
\succ $) if $\mu \left( w\right) \succeq _{w}\emptyset $, for all $w\in W$
and $\mu \left( f\right) =Ch\left( \mu \left( f\right) ,\succ _{f}\right) $
for all $f\in F$.
\par
Note, $\mu \left( w\right) \succeq _{w}\emptyset $ is equivalent to $\mu
\left( w\right) =Ch\left( \mu \left( w\right) ,\succ _{w}\right) .$} Denote
by $IR(\succ )$ the set of individually rational matchings at $\succ $. A
matching $\mu $ is \emph{blocked by a worker-firm pair}\textit{\ }$\left(
w,f\right) $ if $w\notin \mu \left( f\right) $, $w\in Ch\left( \mu \left(
f\right) \cup \left\{ w\right\} ,\succ _{f}\right) $, and \textit{\ }$f\in
Ch(\mu (w)\cup \{f\},\succ _{w})$. A matching $\mu $ is \emph{%
pairwise-stable (stable, for short)} if it is not blocked by any individual
agent or any worker-firm pair.\bigskip

We restrict our attention to a many-to-many matching model where the agents'
preferences satisfy \emph{substitutability} and \emph{Law of Aggregated
Demand }(LAD, for short).\footnote{%
Law of Aggregated Demand was introduce by Alkan (2002), and it called 
\textit{cardinal monotone. }Hatfield and Milgrom (2005) in a model with
Contracts call to this condition Law of Aggregated Demand.}

\begin{definition}
A firm $f$'s preference relation $\succ _{f}$ satisfies substitutability if
for any set $S^{\prime }$ containing workers $w$ and $w^{\prime }$ ($w\neq
w^{\prime }$), if $w\in Ch\left( S^{\prime },\succ _{f}\right) $ then $w\in
Ch\left( S^{\prime }\backslash \{w^{\prime }\},\succ _{f}\right) $.
\end{definition}

That is, if $f$ has substitutable preferences, then if its preferred set of
employees from $S$ includes $w$, so will its preferred set of employees from
any subset of $S$ that still includes $w.$ A preference profile $\succ $ is
substitutable if for each firm $f$, the preference relation $\succ _{f}$
satisfies substitutability.

In an analogous manner, is defined that each worker $w\in W$ has \emph{%
substitutable preferences} over $2^{F}.$

\begin{definition}
The preference $\succ _{f}$ of a firm $f\in F$ satisfies \textbf{law of
aggregated demand} (LAD) if for all $Y\subseteq X\subseteq W$:%
\begin{equation*}
\left\vert Ch(Y,\succ _{f})\right\vert \leq \left\vert Ch(X,\succ
_{f})\right\vert .
\end{equation*}
\end{definition}

A preference profile $(\succ _{f})_{f\in F}$ satisfies \emph{LAD} if for
each firm $f\in F$, the preference relation $\succ _{f},$ satisfies \emph{%
LAD.}

In an analogous manner, is defined that each worker $w\in W$ satisfies \emph{%
law of aggregated demand }$.$\bigskip

Let $\mathbf{M}=(F,W,\succ ,LAD)$ be a specific many-to-many matching model
such that the agents satisfy \emph{substitutability} and \emph{LAD. }We
denote by $\mathcal{S}(\mathbf{M})$ the set of all stable matchings in the
model $\mathbf{M}$.\bigskip

For each $a\in F\cup W,$ we will use the notation $\mu \succeq _{a}\mu
^{\prime }$ which will denote that $\mu \left( a\right) \succ _{a}\mu
^{\prime }\left( a\right) $ or $\mu \left( a\right) =\mu ^{\prime }\left(
a\right) $; $\mu \succ _{a}\mu ^{\prime }$ if $\mu $ $\succeq _{a}\mu
^{\prime }$ and $\mu (a)\neq \mu ^{\prime }(a).$ We denote $\mu \succeq
_{F}\mu ^{\prime }$ if for each $f\in F,$ $\mu \succeq _{f}\mu ^{\prime }.$
We denote $\mu \succ _{F}\mu ^{\prime }$ if $\mu \succeq _{F}\mu ^{\prime }$
and $\mu \neq \mu ^{\prime }.$ Similarly, we denote $\mu \succeq _{W}\mu
^{\prime }$ and $\mu \succ _{W}\mu ^{\prime }.$ Blair (1988) defines a
partial order in the follow way: for each agent $a\in F\cup W,$ we denote
the Blair order by $\mu \succeq _{a}^{B}\mu ^{\prime }$ will denote that $%
\mu (a)=Ch(\mu (a)\cup \mu ^{\prime }\left( a\right) ,\succ _{a})$, and $\mu
\succ _{a}^{B}\mu ^{\prime }$ will denote that $\mu \succeq _{a}^{B}\mu
^{\prime }$ and $\mu \left( a\right) \neq \mu ^{\prime }\left( a\right) $.

\begin{definition}
For a given market $\left( F,W,\succ \right) $, the stable matching $\mu
_{W} $ is called the worker- optimal stable matching if $\mu _{W}\succeq
_{W}\mu $ for every stable matching $\mu $. The firm-optimal stable matching 
$\mu _{F}$ satisfy that $\mu _{F}\succeq _{F}\mu $, for every stable
matching $\mu $.
\end{definition}

\begin{remark}
The firm-optimal stable matching $\mu _{F}$ satisfy $\mu _{F}\succeq
_{F}^{B}\mu $ for every stable matching $\mu $ and inversely. In an
analogous manner, for $\mu _{W}.$
\end{remark}

\begin{remark}
Note that substitutability does not imply LAD and LAD does not imply
substitutability. The preference relation $\succ
_{f}:w_{2},w_{1}w_{3},w_{1},w_{3},\emptyset $\footnote{%
We denote $w_{1}w_{3}=\left\{ w_{1},w_{3}\right\} $}$;$ shows that not all
substitutable preference relations satisfy LAD.
\end{remark}

\section{Matching Game: Preliminaries}

Given the market $(F,W,\succ )$, consider that each agent $a\in F\cup W$ may
replace his/her true preference list, $\succ _{a}$, by any list of
preference $\succ _{a}^{\prime }$, and $\succ ^{\prime }=\left( \succ
_{-a},\succ _{a}^{\prime }\right) $ denote the profile of such lists of
preference, where $\succ _{-a}$ indicates the restriction of $\succ $ to $%
[F\cup W]\backslash \{a\}$. Once $\succ ^{\prime }$ is selected, $W,F$, and $%
\succ ^{\prime }$ are used as \textquotedblleft input\textquotedblright\ for
some algorithm that yields a stable matching for the market $(F,W,\succ
^{\prime })$, as a final \textquotedblleft output\textquotedblright . This
procedure is described by a function $h$ that we call stable matching rule.
For each profile of preferences $\succ ^{\prime }$, $h(\succ ^{\prime })$ is
the stable matching for $(F,W,\succ ^{\prime })$, which is selected by $h.$

Two special stable matching rules are the worker-optimal stable matching
rule $h_{W}$ and the firm-optimal stable matching rule $h_{F}$. Under the
first one (respectively second one) the participants are assigned in
accordance with the worker-optimal (respectively firm-optimal) stable
matching for $\left( F,W,\succ ^{\prime }\right) $. If $\succ ^{\prime }$ is
a profile of preferences, we denote by $h_{W}(\succ ^{\prime })=\mu
_{W}\left( \succ ^{\prime }\right) $ the worker-optimal stable matching for $%
\left( F,W,\succ ^{\prime }\right) $ and by $h_{F}(\succ ^{\prime })=\mu
_{F}\left( \succ ^{\prime }\right) $ the firm-optimal stable matching for $%
\left( F,W,\succ ^{\prime }\right) $.

The adoption of a stable matching rule $h$ for a given market $\left(
F,W,\succ \right) $ induces a strategic game where the set of players is the
set of agents $F\cup W$; a strategy of agent $a\in F\cup W$ is any list of
preferences $\succ _{a}^{\prime }$; the outcome function is determined by $h$
and the true preferences of the players, called true or sincere strategies,
are given by $\succ $. We refer to this game as the matching game induced by 
$h$, and we denote by $\left( F\cup W,\sqsupset ,h,\succ \right) $ where, $%
F\cup W$ is the set of agents, $\sqsupset $ is the set of preference profile
such that all $\succ ^{\prime }\in \sqsupset $ is substitutable, $h$ is the
matching stable rule and $\succ $ is the true or sincere profile of
strategy. We denote by $\left( F\cup W,\sqsupset ,h,\succ ,LAD\right) $ the
matching game where for all $\succ ^{\prime }\in \sqsupset ,$ and $\succ $
are substitutable and $LAD$.

In a many-to-many matching model, the following propositions will be needed
for the proofs of our results.

\begin{proposition}
\label{Alkan}a) \textbf{(Alkan, 2002)}Consider the matching market with
strict substitutable and LAD preferences. Each agent is matched with the
same number of partners in every stable matching.\newline
b) \textbf{(Klijn and Yazici, 2011)} Consider the matching market with
strict, substitutable and $q_{X}$-separable preferences. Then, if an agent
does not fill its quota at some stable matching, it is matched to the same
set of agents at every stable matching.
\end{proposition}

\section{Manipulability Property}

We assume matching model $\left( F,W,\succ \right) $ is substitutable.

Given the matching model $\left( F,W,\succ \right) $, $f\in F$, and $%
T\subseteq W$ satisfying $T=Ch(T,\succ _{f})$ we define $\succ _{f}|_{T}$
the preference $\succ _{f}$ restricted to $T$ as, for all $S^{\prime
}\subseteq W$,

\begin{enumerate}
\item[i.] $S^{\prime }\nsubseteq T$ then $\emptyset \succ _{f}|_{T}$ $%
S^{\prime }$,

\item[ii.] $S^{\prime }\subseteq T$, and $S^{\prime }\succ _{f}$ $\emptyset $
if and only if $S^{\prime }\succ _{f}|_{T}$ $\emptyset $, and

\item[iii.] $S^{\prime },S^{\prime \prime }\subseteq T$ and $S^{\prime
}\succ _{f}S^{\prime \prime }$ if and only if $S^{\prime }\succ
_{f}|_{T}S^{\prime \prime }$.
\end{enumerate}

In an analogous manner, is defined $\succ _{w}|_{T}$ for a worker $w\in W$
and $T\subseteq F.$

\begin{definition}
\label{FMP}Let $\left( F\cup W,\mathcal{\sqsupset },h,\succ \right) $ be a
matching game, we say that $a\in F\cup W$ satisfies \textbf{a manipulability
property} if $h\left( \succ \right) \left( a\right) \neq \mu _{X}\left(
\succ \right) \left( a\right) $ (where, either $X=F,$ if $a\in F$ or $X=W,$
if $a\in W$)$,$ then there exists $\succ _{a}^{\prime }$ such that $h\left(
\succ _{-a},\succ _{a}^{\prime }\right) \succ _{a}h\left( \succ \right) .$
\end{definition}

\begin{definition}
\label{FMPB} Let $\left( F\cup W,\mathcal{\sqsupset },h,\succ \right) $ be a
matching game, we say that $a\in F\cup W$ satisfies \textbf{a manipulability
property with Blair order} if $h\left( \succ \right) \left( a\right) \neq
\mu _{X}\left( \succ \right) \left( a\right) $ (where, either $X=F,$ if $%
a\in F$ or $X=W,$ if $a\in W$)$,$ then there exists $\succ _{a}^{\prime },$
such that $h\left( \succ _{-a},\succ _{a}^{\prime }\right) (a)\succ
_{a}^{B}h\left( \succ \right) (a).$
\end{definition}

The following example show in a many-to-one model with substitutable and $q$%
-separable preferences profile, an agent can manipule and the result of the
stable rule $h$ is a non stable matching in the true profile
strategy.\bigskip

\noindent \textbf{Example\qquad }Let $\left( F,W,\succ \right) $ be where $%
F=\{f_{1},f_{2},f_{3}\}$, $w=\{w_{1},w_{2},w_{3},w_{4}\}$, such that the
agents preferences satisfy substitutability and $q$-separability\emph{. }The
preferences profile $\succ $ is given by;

\begin{equation*}
\begin{tabular}{ll}
$\succ
_{f_{1}}:w_{2}w_{3},w_{2}w_{4},w_{1}w_{3},w_{1}w_{2},w_{1}w_{4},w_{3}w_{4},$
& $\succ _{w_{1}}:f_{1},f_{3},f_{2},$ \\ 
\ \ \ \ \ $w_{1},w_{2},w_{3},w_{4}$ & $\succ _{w_{2}}:f_{2},f_{1},$ \\ 
$\succ _{f_{2}}:w_{1},w_{2}$ & $\succ _{w_{3}}=\succ _{w_{4}}:f_{1},f_{3}.$
\\ 
$\succ _{f_{3}}:w_{4},w_{1}$ & 
\end{tabular}%
\end{equation*}

We have that 
\begin{equation*}
\begin{array}{rccc}
& f_{1} & f_{2} & f_{3} \\ 
\mu _{F}\left( \succ \right) : & w_{2}w_{3} & w_{1} & w_{4} \\ 
\mu _{W}\left( \succ \right) : & w_{1}w_{3} & w_{2} & w_{4}%
\end{array}%
\end{equation*}

Consider that worker $w_{1}$ reject firms $f_{1}$ and $f_{2}$ i.e., $\succ
_{w_{1}}^{\prime }:f_{3},$ and $\succ ^{\prime }=(\succ _{-w_{1}},\succ
_{w_{1}}^{\prime })=(\succ _{w_{1}}^{\prime },\succ _{w_{2}},\succ
_{w_{3}},\succ _{w_{4}},\succ _{f_{1}},\succ _{f_{2}},\succ _{f_{3}})$ then $%
h_{F}\left( \succ ^{\prime }\right) =\mu _{F}^{\prime }$ and we have that%
\begin{equation*}
\begin{array}{rccc}
& f_{1} & f_{2} & f_{3} \\ 
h_{F}(\succ ^{\prime })=\mu _{F}^{\prime }: & w_{3}w_{4} & w_{2} & w_{1}.%
\end{array}%
\end{equation*}

We note that $\mu _{F}^{\prime }\notin S\left( \succ \right) $ because the
pair $(f_{1},w_{1})$ is a blocking pair.

Hence, $h_{F}\left( \succ \right) \neq \mu _{W}$ and $w_{1}$ manipule with $%
\succ _{w_{1}}^{\prime }:f_{3},$ since $h_{F}\left( \succ ^{\prime }\right)
(w_{1})\succ _{w_{1}}h_{F}\left( \succ \right) (w_{1})$ and $h_{F}\left(
\succ ^{\prime }\right) =\mu _{F}^{\prime }\notin S\left( \succ \right) .$ $%
\hfill \square $\bigskip

We are interesting that each agent manipule and the result of $h$ is a
stable matching in the true preference profile. Then, for each $a\in F\cup W$
we define $\ $%
\begin{equation*}
H_{a}=\left\{ \mu \in S(\succ ):\mu \left( a\right) \succ _{a}^{B}h\left(
\succ \right) \left( a\right) \right\} .
\end{equation*}

\begin{remark}
Note that $H_{f}$ $\left( H_{w}\right) $ is non empty because $\mu _{F}\in
H_{f}$ $\left( \mu _{W}\in H_{w},\text{ respectively}\right) .$
\end{remark}

\begin{lemma}
\label{Hf}Let $f\in F$ and $\mu \in H_{f}$. If $\succ _{f}^{\prime \prime
}=\succ _{f}\mid _{\mu (f),}$ then $\mu \in S(\succ ^{\prime \prime })$
where $\succ ^{\prime \prime }=\left( \succ _{-f},\succ _{f}^{\prime \prime
}\right) .$
\end{lemma}

\noindent \textbf{Proof.} \qquad First, we show $\mu \in IR(\succ ^{\prime
\prime }).$ For all $a\in F\cup W,$ if $a\neq f$ we have $\succ _{a}^{\prime
\prime }=\succ _{a}.$ Thus, $Ch(\mu (a),\succ _{a}^{\prime \prime })=Ch(\mu
(a),\succ _{a})=\mu (a).$

For $f,$ we have $\succ _{f}^{\prime \prime }=\succ _{f}|_{%
{\mu}%
\left( f\right) }$, then $Ch(\mu (f),\succ _{f}^{\prime \prime })=\mu (f).$
In effect, if $Ch(\mu (f),\succ _{f}^{\prime \prime })\neq \mu (f),$ hence $%
Ch(\mu (f),\succ _{f}^{\prime \prime })=T$ such that $T\subseteq \mu (f).$
In particular, $T\succ _{f}^{\prime \prime }\mu (f),$ if and only if, by
definition of $\succ _{f}^{\prime \prime },$ $T\succ _{f}\mu (f).$ Thus, $%
\mu (f)\neq Ch(\mu (f),\succ _{f}),$ that is, $\mu \notin IR(\succ ).$ But,
this contradicts that \ $\mu \in S(\succ ).$

Second, there is not exist blocking pair. If there exists $(\overline{f},w)$
blocks $\mu ,$ i.e., $\overline{f}\in Ch(\mu (w)\cup \overline{f},\succ
_{w}^{\prime \prime })$ and $w\in Ch(\mu (\overline{f})\cup w,\succ _{%
\overline{f}}^{\prime \prime }).$

Note that, $\succ _{w}^{\prime \prime }=\succ _{w}$ and $\succ _{\overline{f}%
}^{\prime \prime }=\succ _{\overline{f}}$ if $\overline{f}\neq f.$ Then, if $%
\overline{f}\neq f$ we have that $\mu \notin S(\succ ).$

If $\overline{f}=f,$ $Ch(\mu (f)\cup w,\succ _{f}^{\prime \prime })=\mu (f),$
and this contradicts that $w\in Ch(\mu (f)\cup w,\succ _{f}^{\prime \prime
}) $.

Finally, $\mu \in S(\succ ^{\prime \prime })$.$\hfill \blacksquare
\smallskip $

\begin{proposition}
\label{mu=h}Let $f\in F$, $\mu \in H_{f}$ and $\succ _{f}^{\prime \prime
}=\succ _{f}\mid _{\mu (f),}$then $\mu (f)=h(\succ ^{\prime \prime })(f)$
where $\succ ^{\prime \prime }=\left( \succ _{-f},\succ _{f}^{\prime \prime
}\right) .$
\end{proposition}

\noindent \textbf{Proof.} \qquad By Lemma \ref{Hf}, $\mu \in S(\succ
^{\prime \prime })$ and we have that $h(\succ ^{\prime \prime })\in S(\succ
^{\prime \prime }).$ Hence by definition of order $\succ _{f}^{\prime \prime
}$ we have that $h(\succ ^{\prime \prime })(f)\subseteq \mu (f).$ In effect,
if $h(\succ ^{\prime \prime })(f)\varsubsetneq \mu (f),$ then by definition
of order $\succ _{f}^{\prime \prime },$ we have that $\emptyset \succ
_{f}^{\prime \prime }h(\succ ^{\prime \prime })(f).$ But this contradicts
that \ $h(\succ ^{\prime \prime })\in S(\succ ).$ Then, by Proposition \ref%
{Alkan} (a), $\left\vert h(\succ ^{\prime \prime })(f)\right\vert
=\left\vert \mu (f)\right\vert .$ Thus, $h(\succ ^{\prime \prime })(f)=\mu
(f)$.$\hfill \blacksquare \smallskip $

\begin{proposition}
\label{ifandonlyif}An agent $a\in F\cup W$ satisfies manipulability property
with Blair order, if and only if satisfies manipulability property.
\end{proposition}

\noindent \textbf{Proof.} \qquad In one direction, without lost of
generality we assume that $a=f.$ If $f\in F$ satisfies firm manipulability
property with Blair order, following Definition \ref{FMPB}, there exists $%
\succ _{f}^{\prime }$ such that $h\left( \succ \right) \left( f\right) \neq
\mu _{F}\left( \succ \right) \left( f\right) $ and $h\left( \succ
_{-f},\succ _{f}^{\prime }\right) \succ _{f}^{B}h\left( \succ \right) .$ We
denote $\succ ^{\prime }=\left( \succ _{-f},\succ _{f}^{\prime }\right) ,$
and thus, 
\begin{equation*}
h\left( \succ ^{\prime }\right) \left( f\right) =Ch(h\left( \succ ^{\prime
}\right) \left( f\right) \cup h\left( \succ \right) \left( f\right) ,\succ
_{f})
\end{equation*}%
and $h\left( \succ ^{\prime }\right) (f)\neq h\left( \succ \right) (f).$ We
define $T=h\left( \succ ^{\prime }\right) \left( f\right) \cup h\left( \succ
\right) \left( f\right) ,$ then by definition of choice set we have that $%
h\left( \succ ^{\prime }\right) \left( f\right) \succ _{f}S$ for all $%
S\subset T.$ We note that, in particular, $h\left( \succ \right) \left(
f\right) \subset T.$ Thus, $h\left( \succ ^{\prime }\right) \left( f\right)
\succ _{f}h\left( \succ \right) \left( f\right) $ and $h\left( \succ
^{\prime }\right) (f)\neq h\left( \succ \right) (f).$ Finally, following
Definition \ref{FMP} we have that $f$ satisfies firm manipulability property.

In other direction, without lost of generality we assume that $a=f.$ If $%
f\in F$ satisfies firm manipulability property, following Definition \ref%
{FMP}, there exists $\succ _{f}^{\prime }$ such that $h\left( \succ \right)
\left( f\right) \neq \mu _{F}\left( \succ \right) \left( f\right) $ and $%
h\left( \succ _{-f},\succ _{f}^{\prime }\right) \succ _{f}h\left( \succ
\right) .$

Let $\mu \in H_{f}$ and defined $\succ _{f}^{\prime \prime }=\succ _{f}\mid
_{\mu (f)},$ with $\succ ^{\prime \prime }=(\succ _{-f},\succ _{f}^{\prime
\prime })$ the profile strategy, then by Lemma \ref{Hf}, we have that $\mu
\in S(\succ ^{\prime \prime }).$ Thus, $\mu \in S(\succ ^{\prime \prime })$
and $h(\succ ^{\prime \prime })\in S(\succ ^{\prime \prime }),$ hence by
Proposition \ref{mu=h}, we have that $h(\succ ^{\prime \prime })(f)=\mu (f).$

Since $\mu \in H_{f},$ we have that $\mu \left( f\right) \succ
_{f}^{B}h\left( \succ \right) \left( f\right) $ and $h(\succ ^{\prime \prime
})(f)=\mu (f).$ That is, we prove that if $h\left( \succ \right) \left(
f\right) \neq \mu _{F}\left( \succ \right) \left( f\right) $ then there
exists $\succ _{f}^{\prime \prime }$ such that $h(\succ ^{\prime \prime
})(f)\succ _{f}^{B}h\left( \succ \right) \left( f\right) ,$ i.e., $f$
satisfies manipulability property with Blair order.$\hfill \blacksquare $

\bigskip

We present the main result:

\begin{theorem}[General Manipulability Theorem]
\label{Theorem-1}Let $\left( F\cup W,\mathcal{\sqsupset },h,\succ \right) $
be a matching game with substitutable and LAD preference profile, then for
all $a\in F\cup W$ the manipulability property with Blair order holds.
\end{theorem}

\noindent \textbf{Proof.} Without lost of generality we assume that $a=f.$
Let $f$ be a firm satisfies $h\left( \succ \right) (f)\neq \mu _{F}\left(
\succ \right) \left( f\right) .$ Let $\mu \in H_{f}$ and defined $\succ
_{f}^{\prime \prime }=\succ _{f}\mid _{\mu (f)},$ with $\succ ^{\prime
\prime }=(\succ _{-f},\succ _{f}^{\prime \prime })$ the profile strategy,
then by Lemma \ref{Hf}, we have that $\mu \in S(\succ ^{\prime \prime }).$
Thus, $\mu \in S(\succ ^{\prime \prime })$ and $h(\succ ^{\prime \prime
})\in S(\succ ^{\prime \prime }),$ hence by Proposition \ref{mu=h}, we have
that $h(\succ ^{\prime \prime })(f)=\mu (f).$

Since $\mu \in H_{f},$ we have that $\mu \left( f\right) \succ
_{f}^{B}h\left( \succ \right) \left( f\right) $ and $h(\succ ^{\prime \prime
})(f)=\mu (f).$ That is, we prove that there exists $\succ _{f}^{\prime
\prime }$ such that $h(\succ ^{\prime \prime })(f)\succ _{f}^{B}h\left(
\succ \right) \left( f\right) ,$ i.e., $f$ satisfies manipulability property
with Blair order. We note that, a particular case is $\mu =\mu _{F}$ and the
strategy will be $\mu _{F}.$

$\hfill \blacksquare \smallskip $

Following corollary is consequence of the Proposition \ref{ifandonlyif} and
of the previous Theorem:

\begin{corollary}
Let $\left( F\cup W,\mathcal{\sqsupset },h,\succ \right) $ be a matching
game with substitutable and LAD preference profile, then for all $a\in F\cup
W$ the manipulability property holds.
\end{corollary}

\begin{remark}
As consequence of General Manipulability Theorem in a many-to-many matching
game with substitutable and LAD preference profile, the General
Impossibility Theorem and the Impossibility Theorem holds.
\end{remark}

\section{General Manipulability Theorem with substitutable preferences}

In this Section, in the many-to-one model with substitutable preferences we
present two examples where show the General Manipulability Theorem \emph{is
false}. We presents the following two examples due to this model does not
symmetric. Sotomayor (2012) proved the General Manipulability Theorem for
the case many-to-one with responsive preferences. Thus, both examples show
Sotomayor (2012) results' can not be generalized to substitutable preference
profile.\footnote{%
Remember, substitutability is a weaker condition than responsiveness.} The
Example 1, shows the firms can not manipulated, and Example 2, shows the
workers can not manipulated.\bigskip 

\noindent \textbf{Example 1\qquad } Let $\left( F,W,\succ \right) $ be where 
$F=\{f_{1},f_{2},f_{3}\}$, $w=\{w_{1},w_{2},w_{3},w_{4}\}$, and the
preferences profile $\succ $ is given by;%
\begin{equation*}
\begin{array}{lcrl}
\succ _{f_{1}}:w_{1}w_{2},w_{1},w_{2},w_{3}w_{4},w_{3},w_{4}, &  & \succ
_{w_{1}}=\succ _{w_{2}}: & f_{2},f_{3},f_{1}, \\ 
\succ _{f_{2}}:w_{3},w_{1}w_{4},w_{4},w_{1}w_{2},w_{1},w_{2}, &  & \succ
_{w_{3}}: & f_{1},f_{3},f_{2}, \\ 
\succ _{f_{3}}:w_{4},w_{2}w_{3},w_{1}w_{2},w_{3},w_{1},w_{2}, &  & \succ
_{w_{4}}: & f_{1},f_{2},f_{3}.%
\end{array}%
\end{equation*}%
Let $h_{W}$ be a optimal-worker stable matching rule, i.e., for all
preference $\succ ^{\prime }$, $h_{W}(\succ ^{\prime })=$ $\mu _{W}\left(
\succ ^{\prime }\right) $.

Note that $\succ $ is substitutable but does not satisfy LAD. For example,
consider the preference $\succ _{f_{1}},$ and $A=\left\{
w_{2},w_{3},w_{4}\right\} ,$ and $B=\left\{ w_{3},w_{4}\right\} .$ Thus, $%
B\subset A$, $Ch(A,\succ _{f_{1}})=\{w_{2}\}$ and $Ch(B,\succ
_{f_{1}})=\{w_{3},w_{4}\}.$ Hence, $\left\vert Ch(B,\succ
_{f_{1}})\right\vert >\left\vert Ch(A,\succ _{f_{1}})\right\vert ,$ that is,
the preferences $\succ _{f_{1}}$does not satisfy $LAD$. In the same way, the
preferences $\succ _{f_{2}}$ and $\succ _{f_{3}}$do not satisfy $LAD$ .

We have that%
\begin{equation*}
\begin{array}{rccc}
& f_{1} & f_{2} & f_{3} \\ 
h_{W}(\succ )=\mu _{W}\left( \succ \right) : & w_{3}w_{4} & w_{1}w_{2} & 
\emptyset \\ 
\mu _{F}\left( \succ \right) : & w_{1}w_{2} & w_{3} & w_{4}%
\end{array}%
\end{equation*}%
Consider that firm $f_{1}$ reject workers $w_{3}$ and $w_{4}$ i.e., $\succ
_{f_{1}}^{\prime }=\succ _{f_{1}}|_{\mu _{F}\left( \succ \right) \left(
f_{1}\right) }:w_{1}w_{2},w_{1},w_{2}$ and $\succ ^{\prime }=(\succ
_{-f_{1}},\succ _{f_{1}}^{\prime })$ then $h_{W}(\succ ^{\prime })=\mu
_{W}\left( \succ ^{\prime }\right) $ and we have that%
\begin{equation*}
\begin{array}{rccc}
& f_{1} & f_{2} & f_{3} \\ 
h_{W}(\succ ^{\prime })=\mu _{W}\left( \succ ^{\prime }\right) : & \emptyset
& w_{1}w_{4} & w_{2}w_{3}.%
\end{array}%
\end{equation*}%
Thus, $h_{W}(\succ ^{\prime })(f_{1})=\emptyset $ and $h_{W}(\succ
)(f_{1})=\left\{ w_{3},w_{4}\right\} .$ Hence, $h_{W}(\succ )(f_{1})\succ
_{f_{1}}h_{W}(\succ ^{\prime })(f_{1}),$ in particular $h_{W}(\succ
)(f_{1})\succ _{f_{1}}^{B}h_{W}(\succ ^{\prime })(f_{1}).$ Therefore, firm $%
f_{1}$ has not incentive to misrepresent its preference.

Consider that firm $f_{2}$ reject any set that contain worker $w_{1},w_{2}$
and $w_{4},$ i.e., $\succ _{f_{2}}^{\prime }:w_{3}$ and $\succ ^{\prime
}=(\succ _{-f_{2}},\succ _{f_{2}}^{\prime })$ then $h_{W}(\succ ^{\prime })=$
$\mu _{W}\left( \succ ^{\prime }\right) $ we have that%
\begin{equation*}
\begin{array}{rccc}
& f_{1} & f_{2} & f_{3} \\ 
h_{W}(\succ ^{\prime })=\mu _{W}\left( \succ ^{\prime }\right) : & w_{3}w_{4}
& \emptyset & w_{1}w_{2}.%
\end{array}%
\end{equation*}%
Thus, $h_{W}(\succ ^{\prime })(f_{2})=\emptyset $ and $h_{W}(\succ
)(f_{2})=\left\{ w_{1},w_{2}\right\} .$ Hence, $h_{W}(\succ )(f_{2})\succ
_{f_{2}}h_{W}(\succ ^{\prime })(f_{2}),$ in particular $h_{W}(\succ
)(f_{2})\succ _{f_{2}}^{B}h_{W}(\succ ^{\prime })(f_{2}).$ Therefore, firm $%
f_{2}$ has not incentive to misrepresent its preference.

Consider that firm $f_{3}$ reject any set that contain worker $w_{1},w_{2}$
and $w_{3},$ i.e., $\succ _{f_{3}}^{\prime }:w_{4}$ and $\succ ^{\prime
}=(\succ _{-f_{3}},\succ _{f_{3}}^{\prime })$ then $h_{W}(\succ ^{\prime })=$
$\mu _{W}\left( \succ ^{\prime }\right) $ we have that%
\begin{equation*}
\begin{array}{rccc}
& f_{1} & f_{2} & f_{3} \\ 
h_{W}(\succ ^{\prime })=\mu _{W}\left( \succ ^{\prime }\right) : & w_{3}w_{4}
& w_{1}w_{2} & \emptyset .%
\end{array}%
\end{equation*}%
Thus, $h_{W}(\succ ^{\prime })(f_{3})=\emptyset $ and $h_{W}(\succ
)(f_{3})=\emptyset .$ Hence, $h_{W}(\succ )(f_{3})=h_{W}(\succ ^{\prime
})(f_{3}).$ Therefore, firm $f_{3}$ has not incentive to misrepresent its
preference. $\hfill \square $\bigskip

\noindent \textbf{Example 2\qquad }Let $\left( F,W,\succ \right) $ be where $%
F=\{f_{1},f_{2}\}$, $W=\{w_{1},w_{2},w_{3},w_{4}\}$, and the preferences
profile $\succ $ is given by;%
\begin{equation*}
\begin{array}{ccrlc}
\succ _{f_{1}}:w_{1}w_{2},w_{1},w_{2},w_{3}w_{4},w_{3},w_{4}, &  & \succ
_{w_{1}}=\succ _{w_{2}}: & f_{2},f_{1}, &  \\ 
\succ _{f_{2}}:w_{3}w_{4},w_{3},w_{4},w_{1}w_{2},w_{1},w_{2}, &  & \succ
_{w_{3}}=\succ _{w_{4}}: & f_{1},f_{2}. & 
\end{array}%
\end{equation*}

Let $h_{F}$ be a firm-optimal stable matching rule.

Note that $\succ $ is substitutable but does not satisfy LAD. For example,
consider the preference $\succ _{f_{1}},$ and $A=\left\{
w_{2},w_{3},w_{4}\right\} ,$ and $B=\left\{ w_{3},w_{4}\right\} .$ Thus, $%
B\subset A$, $Ch(A,\succ _{f_{1}})=\{w_{2}\}$ and $Ch(B,\succ
_{f_{1}})=\{w_{3},w_{4}\}.$ Hence, $\left\vert Ch(B,\succ
_{f_{1}})\right\vert >\left\vert Ch(A,\succ _{f_{1}})\right\vert ,$ that is,
the preferences $\succ _{f_{1}}$does not satisfy LAD. In the same way, the
preferences $\succ _{f_{2}}$ does not satisfy LAD.

We have that%
\begin{equation*}
\begin{array}{rcccc}
& w_{1} & w_{2} & w_{3} & w_{4} \\ 
\mu _{W}\left( \succ \right) : & f_{2} & f_{2} & f_{1} & f_{1} \\ 
h_{F}(\succ )=\mu _{F}\left( \succ \right) : & f_{1} & f_{1} & f_{2} & f_{2}%
\end{array}%
\end{equation*}

Then $h_{F}(\succ )\neq \mu _{W}$. Consider $\succ ^{\prime }=\left( \succ
_{-w_{1}},\succ _{w_{1}}^{\prime }\right) =(\succ _{w_{1}}^{\prime },\succ
_{w_{2}},\succ _{w_{3}},\succ _{w_{4}},\succ _{f_{1}},\succ _{f_{2}})$ then $%
h_{F}(\succ ^{\prime })=\mu _{F}\left( \succ ^{\prime }\right) $ we have
that for any $\succ _{w_{1}}^{\prime }$, $h_{F}(\succ ^{\prime })(w_{3})=%
{\mu}%
_{F}\left( \succ ^{\prime }\right) (w_{3})=f_{2}$. Hence, worker $w_{1}$,
only can get firm $f_{1}$, then he has not incentive to misrepresent its
preference. Similarly worker $w_{2}$, $w_{3},$ and $w_{4}$ have not
incentive to misrepresent its true preference $\succ .\hfill \square
\smallskip $

\section{Concluding remarks}

This paper contributes to literature in proves three important results which
fill a gap in the theory of incentives for the many-to-many model. This
results are responsible for the success that the incentives theory has had
in explaining empirical economic phenomena in the many-to-many matching
models.

\section{Acknowledgements}

We acknowledge financial support from the Universidad Nacional de San Luis,
through grant 03-2016, and from Consejo Nacional de Investigaciones Cient%
\'{\i}ficas y T\'{e}cnicas (CONICET) through grant PIP 112-200801-00655, and
from Agencia Nacional de Promoci\'{o}n Cient\'{\i}fica y Tecnol\'{o}gica
through grant PICT 2017-2355.

\section*{References}

\begin{enumerate}
\item Alkan, A. (2002). \textquotedblleft A Class of Multipartner Matching
Models with a Strong Lattice Structure,\textquotedblright\ \textit{Economic
Theory}, \textbf{19}: 737-746.

\item Blair, C. (1988). \textquotedblleft The Lattice Structure of the Set
of Stable Matchings with Multiple Partners,\textquotedblright\ \textit{%
Mathematics of Operations Research}, \textbf{13}: 619-628.

\item Dubins and Freedman, (1981). \textquotedblleft Machiavelli and the
Gale-Shapley Algorithm,\textquotedblright\ \textit{American Mathematical
Monthly}, \textbf{88}: 485-494.

\item Gale, D. and Shapley, L. (1962). \textquotedblleft College Admissions
and Stability of Marriage,\textquotedblright\ \textit{American Mathematical
Monthly}, \textbf{69}: 9-15.

\item Gale, D. and Sotomayor, M. (1985). \textquotedblleft Some Remarks on
the Stable Marriage Problem,\textquotedblright \textit{\ Discrete Applied
Mathematics}, \textbf{11}: 223-232.

\item Hatfield, J. W. and Kojima, F. (2010). \textquotedblleft Substitutes
and Stability for Matching with Contracts,\textquotedblright\ Journal of
Economic Theory, \textbf{145}: 1704-1723.

\item Hatfield, J.W. and Milgrom, P. (2005). \textquotedblleft Matching with
Contracts,\textquotedblright\ \textit{American Economic Review,} \textbf{95}%
: 913-935.

\item Kelso, A. and Crawford, V. (1982). \textquotedblleft Job Matching,
Coalition Formation, and Gross Substitutes,\textquotedblright\ \textit{%
Econometrica}, \textbf{50}: 1483-1504.

\item Klijn, F. and Yazici, A. (2011). \textquotedblleft A Many-to-Many
Hospital Theorem\textquotedblright , Barcelona Graduate School of Economics.
Working Paper Series N%
${{}^o}$
567 (2014).

\item Martinez, R., Mass\'{o}, J., Neme, A. and Oviedo, J. (2004).
\textquotedblleft On Group Strategy-Proof Mechanisms for a Many-to-One
Matching Model,\textquotedblright\ International Journal of Game Theory%
\textit{, }\textbf{33}: 115-128.

\item Roth, A. (1982). \textquotedblleft The Economics of Matching:
Stability and Incentives \textquotedblright , \textit{Mathematics of
Operations Research}, \textbf{7}: 617-628.

\item Roth, A. (1984). \textquotedblleft The Evolution of the Labor Market
for Medical Interns and Residents: A Case Study in Game
Theory,\textquotedblright\ \textit{Journal of Political Economy}, \textbf{92}%
: 991-1016.

\item Roth, A. (1985). \textquotedblleft The College Admissions Problem is
not Equivalent to the Marriage Problem,\textquotedblright\ \textit{Journal
of Economic Theory}, \textbf{36}: 277-288.

\item Roth, A. (1986). \textquotedblleft On the Allocation of Residents to
Rural Hospitals: A General Property of Two-Sided Matching
Markets,\textquotedblright\ \textit{Econometrica }\textbf{54}: 425-427.

\item Roth, A. and Sotomayor, M. (1990).\textit{\ Two-Sided Matching: A
Study in Game-Theoretic Modeling and Analysis.} Econometric Society
Monographs, Vol. 18. Cambridge University Press, Cambridge, England.

\item Sotomayor, M. (1996). \textquotedblleft Admission mechanisms of
students to colleges. A game-theoretic modeling and
analysis\textquotedblright , Brazilian Review of Econometrics \textbf{16}:
25-63.

\item Sotomayor, M. (2012). \textquotedblleft A Further Note on the College
Admission Game,\textquotedblright\ \textit{International Journal Game Theory}%
, \textbf{41}: 179-193.
\end{enumerate}

\end{document}